\documentclass[a4paper,fleqn]{cas-dc}
\usepackage[numbers]{natbib}
\usepackage{graphicx}
\usepackage{booktabs}
\usepackage{multirow}

\def\tsc#1{\csdef{#1}{\textsc{\lowercase{#1}}\xspace}}
\tsc{WGM}
\tsc{QE}

\begin{document}
\let\WriteBookmarks\relax
\let\printorcid\relax 

\shorttitle{} 
\shortauthors{RenJie Zheng et al.}

\title[mode = title]{Accelerating NBTI Aging Evaluation via Physics-Aware Graph Attention Networks}  

\author[1]{RenJie Zheng}
\author[2]{Hengxi Liu}
\author[2]{ShuJun Gao}
\author[2]{Xinyue Feng}
\author[2]{Hailong You}
\author[2]{Cong Li}
\cormark[1]

\address[1]{School of Microelectronics, Xidian University, Xi’an 710071, China; } 
\cortext[1]{Corresponding author:Cong Li}  

\begin{abstract}
As semiconductor technology advances to smaller nodes, Negative Bias Temperature Instability (NBTI) under prolonged workloads has emerged as a significant bottleneck constraining reliability-aware Design-Technology Co-Optimization (DTCO). Conventional TCAD simulations incur prohibitive computational overhead when evaluating device aging characteristics, making it difficult to satisfy the demand for efficient iterative design cycles. To address this challenge, this paper proposes an aging evaluation framework based on a physics-aware graph attention network (Physics-Aware RelGAT). By losslessly mapping unstructured device meshes into attributed graphs, this framework constructs a 45-dimensional device encoding scheme that integrates interface trap distributions and macroscopic electro-thermal stresses, achieving a direct mapping from underlying physical quantities to device degradation characteristics. To overcome the challenge of predicting currents that span multiple orders of magnitude, a dual-end normalization strategy and a log-scale loss function optimization are introduced, ensuring the model possesses high-precision fitting capabilities. Experimental results demonstrate that the model achieves a mean error of only 1.27\% on an independent test set, achieving an acceleration of approximately 17,000 times compared to traditional TCAD simulations. This framework provides a solution for the assessment of circuit reliability in advanced process nodes that successfully balances physical fidelity with industrial-grade efficiency.
\end{abstract}


\maketitle

\begin{figure*}[pos=t] 
    \centering
    \includegraphics[width=\textwidth]{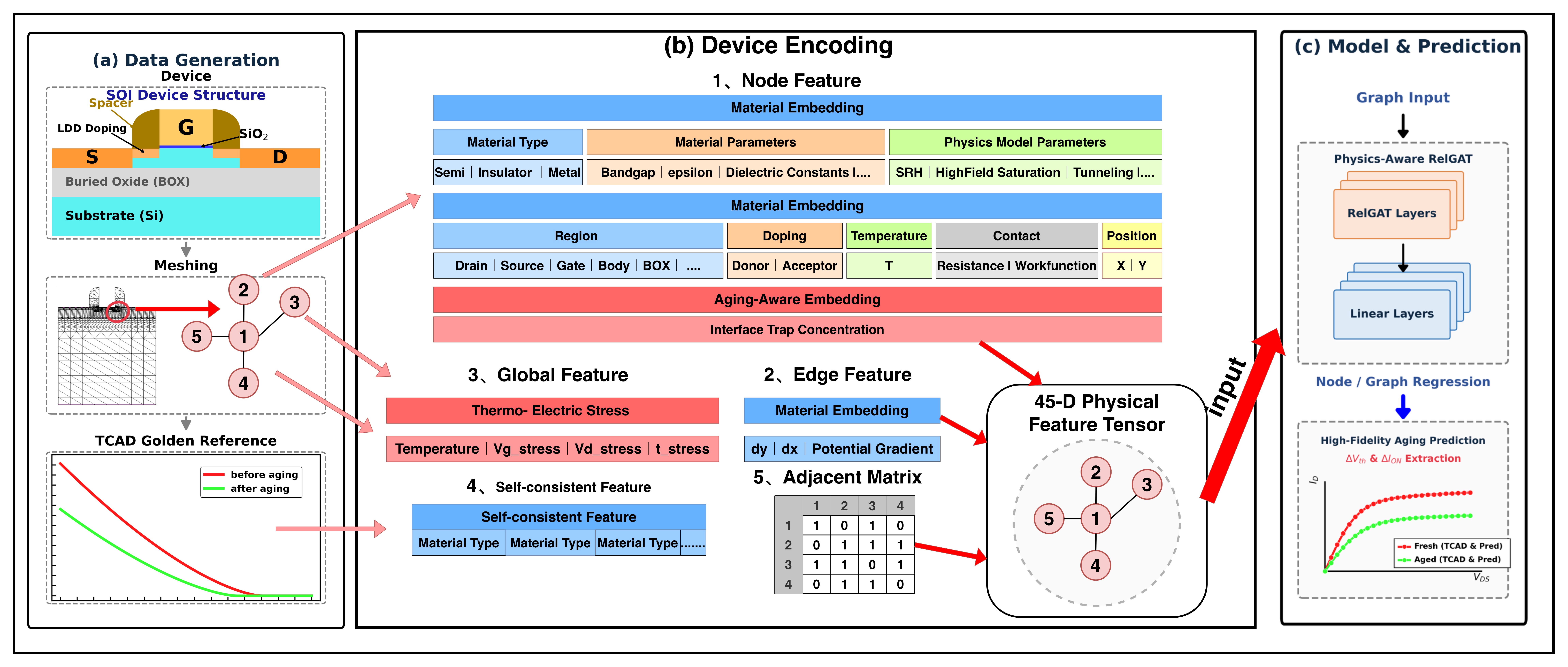}
    \caption{The proposed unified aging-aware graph encoding scheme, illustrating the fusion of microscopic node features and macroscopic time-varying global stress parameters.}
    \label{fig:graph_encoding}
\end{figure*}

\section{Introduction}

As integrated circuits continue to scale towards advanced process nodes, the extreme miniaturization of device structures and exceedingly high internal electric fields have severely exacerbated aging effects such as Negative Bias Temperature Instability (NBTI) and Hot Carrier Injection (HCI). This not only alters the electrical characteristics of microscopic devices, but also emerges as a major challenge that threatens the long-term stability of large-scale digital circuits and their end-of-life (EOL) performance \cite{Mukhopadhyay2023, Chatterjee2023}. To accurately evaluate and mitigate such system-level failures induced by physical degradation during the circuit design phase, while simultaneously avoiding the "over-design" and performance waste caused by traditional conservative estimations, establishing a reliability-oriented "aging-aware Design-Technology Co-Optimization" (DTCO) framework has become a fundamental cornerstone for balancing Power, Performance, Area, and Reliability (PPAR) in digital circuit development \cite{Kelleher2022, Moroz2020}.

Within the Aging-Aware DTCO flow, TCAD simulation serves as a "virtual fab," providing underlying physical fidelity \cite{Novkin2023}. Compared to high-cost silicon prototyping (tape-out tests), it can numerically solve fundamental physical equations to accurately capture internal electric fields and trap dynamics that are difficult to observe directly in experiments \cite{Xue2023}. This high-precision capability, equipped with powerful physical extrapolation, makes it the core physical engine for evaluating the aging characteristics of novel architectures during the technology pathfinding stage \cite{Zhao2019, Tariq2025}.

However, as process nodes advance to 3 nm and below, conventional TCAD simulations face severe computational challenges \cite{Novkin2023, Mohamed2025}. On the one hand, the evolution of device architectures toward complex 3D structures such as FinFET, Gate-All-Around (GAA), and even CFETs has led to a dramatic surge in mesh grid counts \cite{Mohamed2025}. Coupled with the introduction of quantum mechanical correction models and complex non-ideal effects in advanced devices, the difficulty of Newton iteration convergence in single-point simulations has significantly increased \cite{Mohamed2025, Han2021_conf}. In practical engineering applications, this computational bottleneck is particularly pronounced in aging simulations. With the introduction of complex gate structures, microscopic defect networks become extremely intricate, rendering traditional models inadequate for precisely describing the complex dynamic capture and emission behaviors of defects, which in turn makes it difficult to accurately characterize device aging \cite{Xue2023}. Furthermore, to comprehensively evaluate aging characteristics, simulations must iterate across a vast array of bias combinations, temperature ranges, and aging times spanning multiple orders of magnitude \cite{Mukhopadhyay2023, Xue2023}. Under such massive simulation demands, conventional TCAD must execute transient simulations across enormous time scales to accurately capture the dynamic evolution of microscopic defects. The superimposition of an extremely large cross-simulation matrix and a lengthy single-point transient solving process means that precisely evaluating the global aging behavior of a class of advanced devices often requires days or even longer computation times \cite{Novkin2023, Han2021_jour}. When the prolonged duration of a single simulation point is compounded with the massive volume of cross-simulation tasks required for a comprehensive assessment, reliability-aware DTCO incurs exorbitant time costs at the device level. This severe computational bottleneck significantly lengthens the iteration cycle of aging simulations, drastically increasing evaluation overhead in practical engineering applications \cite{Novkin2023, Han2021_jour}.

In recent years, the introduction of deep learning and neural network technologies has provided new solutions to break through the speed bottleneck of TCAD simulations. Researchers have successfully integrated machine learning as surrogate models into device design \cite{Novkin2023}. Initially, researchers attempted to construct purely data-driven surrogate models using Artificial Neural Networks (ANN) to achieve rapid modeling through the numerical mapping of device electrical characteristics \cite{Wei2022_tcad, Wei2022_integ}. Although such methods exhibit excellent fitting precision at specific process nodes, they are fundamentally "black-box" models lacking descriptions of internal physical topology and spatial field distributions. Due to the absence of physical constraints, black-box models often suffer from limited generalization capabilities and poor physical extrapolation, making it difficult to capture the dynamic evolution of microscopic physical quantities. To break the physical barriers of black-box models and introduce structure-awareness, Han et al. utilized Convolutional Neural Networks (CNN) to provide high-precision initial potential guesses for the TCAD solver. While ensuring 100\% physical accuracy, this approach significantly reduced the iterative simulation time of FinFETs by improving the accuracy of the initial solutions \cite{Han2021_conf, Han2021_jour}. However, because the native architecture of CNNs relies heavily on regular structured grids, treating the highly irregular unstructured meshes prevalent in TCAD simulations necessitates forcibly interpolating the device's spatial data into a regular 3D pixel array. This mathematical interpolation inevitably triggers severe geometric distortion and blurs crucial subtle physical features at the device interfaces, resulting in non-negligible accuracy losses.

To overcome the errors generated by forced interpolation when CNNs process unstructured meshes, Jang et al. pioneered the application of Graph Neural Networks (GNN) to device modeling. By directly transforming unstructured meshes into graphs, this approach avoids the accuracy loss introduced by interpolation, achieving the lossless extraction of internal physical boundaries and electrical characteristics \cite{Jang2023}. Building on this, Fan et al. proposed a unified device encoding scheme that embeds material properties and spatial positional relationships into graph representations. Utilizing a Relational Graph Attention Network (RelGAT), they constructed an end-to-end surrogate model, successfully realizing the simulation evaluation of emerging material devices (e.g., carbon nanotubes) through transfer learning based on mature silicon-based devices \cite{Fan2025}. As research has deepened, the application of neural networks has extended to complex physical fields such as electro-thermal coupling. For instance, Mohamed et al. achieved rapid prediction of device self-heating distributions based on CNNs \cite{Mohamed2025}.

Despite the significant phased progress in physical field modeling and simulation acceleration achieved by the aforementioned studies, there remains a research gap in the domain of device aging simulation. First, graph representation works, represented by reference \cite{Fan2025}, are currently limited to static predictions of single-point transistor currents, failing to efficiently characterize high-dimensional transfer characteristic curves across the full voltage sweep range. More importantly, facing the highly time-consuming NBTI aging simulation, there is still no precedent for directly applying deep learning to accelerate its underlying physical simulation. Consequently, the aging-aware DTCO flow remains constrained by the computational shackles of conventional TCAD. Therefore, referring to and adopting the unified device encoding scheme and RelGAT methodology proposed in \cite{Fan2025}, this paper proposes an aging-aware graph encoding scheme, as illustrated in Fig. \ref{fig:graph_encoding}, and conducts the following research focused on aging simulation:

\begin{itemize}
    \item \textbf{Realizing high-fidelity prediction of transfer characteristic curves across the full bias range.} This paper extends the model output to high-dimensional $I_d-V_g$ curve tensors, achieving an end-to-end mapping from unstructured meshes to device external characteristics, comprehensively portraying both subthreshold and ON-state characteristics. Experiments demonstrate that the model achieves an average error of only 1.27\% on an independent test set, with a 95th percentile error below 2.92\%, exhibiting extremely high physical fidelity across the full operational range.
    \item \textbf{Reconstructing the physical data flow to deeply accelerate aging simulations.} Addressing the flaw that traditional graph pooling tends to dilute local critical extrema, this paper extracts the NBTI-induced interface trap charges and injects them as features into the fully connected prediction head. This design accurately captures dynamic degradation trajectories and breaks the computational bottleneck of conventional TCAD. The model's inference time for a single curve is only 3.17 milliseconds, achieving an approximately 17,000$\times$ speedup over TCAD and substantially shortening the DTCO iteration cycle.
    \item \textbf{Introducing dual-end normalization and log-scale strategies to overcome numerical disasters.} Targeting the challenges of immense disparities in input physical features of advanced devices and gradient blocking caused by output currents spanning eight orders of magnitude, this paper implements static scale governance simultaneously at the feature input and label output ends, and introduces a log-scale loss function. This strategy thoroughly eliminates computational distortion in high-dimensional mapping, ensuring robust model convergence across the entire current range.
\end{itemize}

\section{Methodology}

\subsection{Physical Mechanism of NBTI Degradation}

\begin{figure}[pos=ht]
    \centering
    \includegraphics[width=0.9\linewidth]{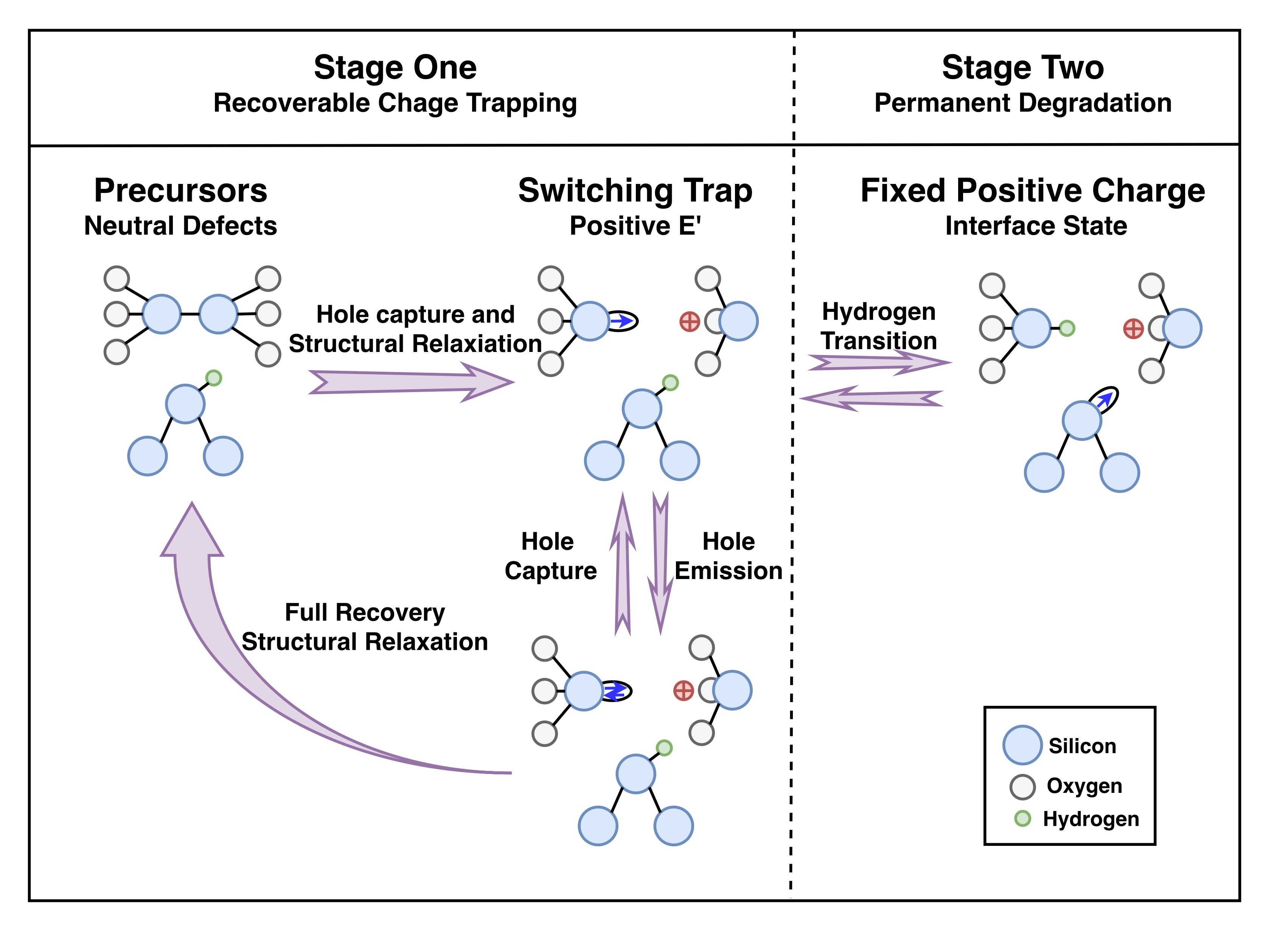}
    \caption{The Two-Stage degradation model for NBTI, illustrating the transition among four physical states during the recoverable charge trapping and permanent degradation stages.}
    \label{fig:nbti_mechanism}
\end{figure}

In the reliability assessment of semiconductor devices, Negative Bias Temperature Instability (NBTI) is the core mechanism causing the performance degradation of PMOS transistors \cite{Stathis2006}. To reflect the physical degradation process of the device, this paper adopts the Two-Stage degradation model built into Sentaurus TCAD \cite{Grasser2009, Goes2009}. This model decouples the NBTI-induced interface defect generation mechanism into two interconnected stages: "recoverable charge trapping" and "permanent interface degradation", involving a total of four physical states, as illustrated in Fig. \ref{fig:nbti_mechanism}.

The first stage encompasses three states (S1-S3): S1 represents the pre-existing Si-Si bond defects (degradation precursors) in the gate oxide layer. Under electro-thermal stress, channel holes tunnel into the oxide and are captured, causing the Si-Si bonds to break and forming positively charged E'-Centers (State S2). If the holes in S2 are released through tunneling, they transition into neutral E'-Centers (State S3), a state that can structurally relax and recover back to S1.

The second stage (State S4) is transformed from S2: the positively charged E'-Centers further trigger the breaking of Si-H bonds at the interface, generating unpassivated silicon dangling bonds. These continuously accumulating silicon dangling bonds (i.e., Interface Traps) constitute the core physical mechanism leading to the permanent and irreversible electrical performance degradation of PMOS under long-term stress.

In transient numerical simulations, the trap density generated over time is coupled into the Poisson's equation and carrier transport equations. Solving these non-linear equations coupled with electro-thermal-defect dynamics significantly increases the convergence difficulty and computational time overhead of the simulation. Therefore, developing an efficient AI surrogate model equipped with physics-awareness becomes an inevitable choice for accelerating reliability-aware Design-Technology Co-Optimization (DTCO).

\subsection{Unified Aging-Aware Graph Encoding Scheme}

To achieve an efficient and generalized mathematical representation of physical fields within complex semiconductor devices, this paper proposes a unified aging-aware graph encoding scheme based on the methodology established by Fan et al. \cite{Fan2025}, as illustrated in Fig. \ref{fig:graph_encoding}. The core of this scheme lies in mapping the unstructured discrete meshes generated by the TCAD simulator into a directed graph $G=(V,E)$. Within this mapping framework, $V$ denotes the set of nodes corresponding directly to the mesh vertices, whereas $E$ represents the set of edges defined by the mesh topological connections. In the construction of the baseline graph embedding, the framework incorporates representations across four distinct dimensions: material-level embeddings utilizing one-hot encoding to distinguish the physical boundaries of heterojunction materials; device-level embeddings providing the exact geometric coordinates $(x,y)$ of the mesh; global embeddings capturing the macroscopic bias voltages and temperatures; and spatial relationship embeddings that extract relative distances and potential gradients between adjacent nodes to emulate finite element interpolation characteristics \cite{Han2021_conf}.

To endow this graph encoding framework with explicit aging-awareness, this paper introduces an aging extension built upon the baseline transfer characteristic curve representation. At the microscopic node feature level, the interface trap density distribution—which constitutes the core physical quantity governing device performance degradation—is directly injected into the node attributes. This enhancement ensures that the graph topology reflects not merely the static geometry, but dynamically captures the microscopic physical origin of device degradation.

Furthermore, the electro-thermal stresses sustained by the device during operation (such as stress temperature $T$, stress gate bias $V_{g\text{\_stress}}$, and stress drain bias $V_{d\text{\_stress}}$) along with the cumulative stress time ($t_{\text{stress}}$) serve as the decisive factors governing its degradation trajectory. Consequently, these parameters are extended into the global features of the graph. Utilizing a global broadcasting mechanism, this aging state vector is concatenated into the feature vector of every single node within the graph. Ultimately, each node is assigned a comprehensive 45-dimensional composite physical feature tensor.

This fusion encoding paradigm, which couples microscopic defect attributes with macroscopic time-varying operational stresses, successfully transforms the device representation from a static topology into an aging-stress-aware dynamic physical system. This multi-dimensional encoding lays a solid foundation for achieving full-range, high-fidelity aging predictions in subsequent stages.

\subsection{Physics-Aware Graph Attention Networks}

\begin{figure*}[pos=ht]
     \centering
    \includegraphics[width=0.8\textwidth]{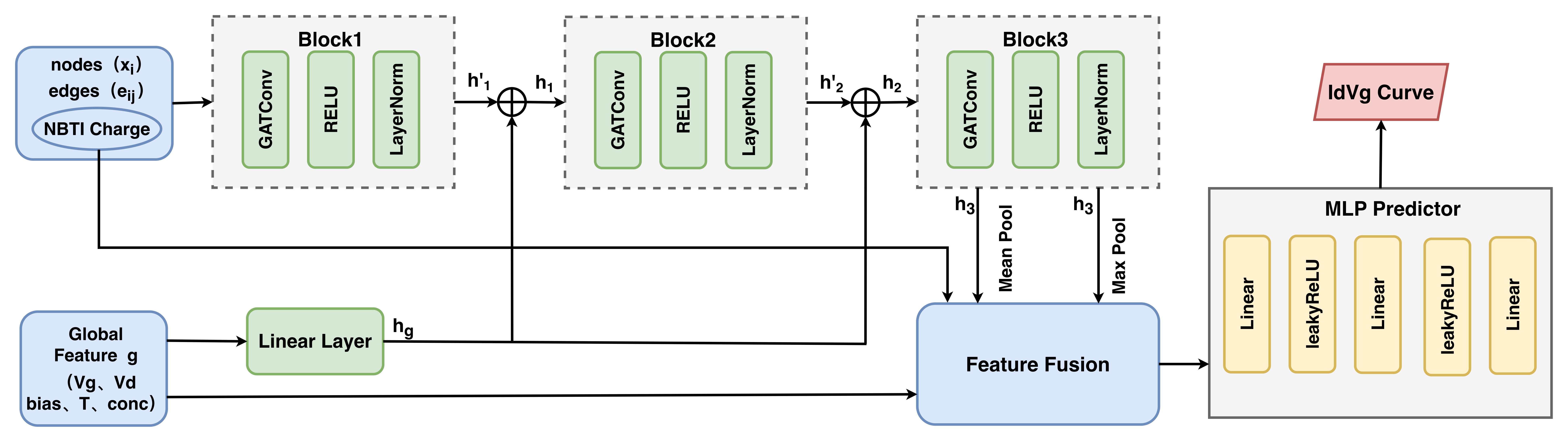}
    \caption{The overall architecture of the proposed Physics-Aware RelGAT model.}
    \label{fig:overall_model}
\end{figure*}

In TCAD numerical simulations, the distribution of electrostatic potential and carrier concentration within the device relies on unstructured meshes. To preserve the topological information of this microscopic physical space, the TCAD simulation mesh of the device is directly mapped into a directed attributed graph $G=(V,E)$. Here, the node set $V$ represents the TCAD mesh nodes, and the edge set $E$ represents the physical connections between nodes. Each node $i$ is assigned a 45-dimensional feature vector $h_i \in \mathbb{R}^{45}$. This 45-dimensional feature constitutes the ``physics-aware domain'' of the model. It encapsulates not only the static device geometry (e.g., $L_{gate}$, $t_{ox}$) and doping profiles (e.g., $N_{sd}$, $N_{ch}$), but also deeply integrates global bias conditions ($V_g$, $V_d$), dynamic aging stress time ($t_{stress}$), and interface defect density. Fig. \ref{fig:overall_model} demonstrates the overall model architecture.

Because different internal regions of the device (such as the channel surface and PN junction depletion regions) contribute with extreme non-linear weight disparities to the overall terminal current, traditional Graph Convolutional Networks (GCN) cannot adaptively capture this spatial anisotropy. Therefore, the core operator of this paper employs the Graph Attention Network (GAT). During the feature aggregation process at the $l$-th layer, the physical coupling strength (attention coefficient) $e_{ij}^{(l)}$ between node $i$ and its neighboring node $j$ is calculated by the following equation:

\begin{equation}
e_{ij}^{(l)}=\mathrm{LeakyReLU}\left(\mathbf{a}^T\left[\mathbf{W}^{(l)}h_i^{(l)}\parallel\mathbf{W}^{(l)}h_j^{(l)}\parallel f_{e,ij}\right]\right)
\label{eq:attention_coeff}
\end{equation}

where $\mathbf{W}^{(l)}$ is a learnable linear transformation weight matrix, $\parallel$ denotes the feature concatenation operation, $\mathbf{a}$ represents the feed-forward neural network weights of the attention mechanism, and $f_{e,ij}$ is the edge feature (including the Euclidean distance between nodes and mesh geometry information). To ensure stable gradient propagation and comparability of weights, a softmax function is applied to normalize the coefficients across all neighboring nodes $\mathcal{N}(i)$ of the central node $i$:

\begin{equation}
\alpha_{ij}^{(l)}=\frac{\exp(e_{ij}^{(l)})}{\sum_{k\in\mathcal{N}(i)}\exp(e_{ik}^{(l)})}
\label{eq:softmax_norm}
\end{equation}

Subsequently, the microscopic features of the current node are updated by aggregating the information from its neighboring nodes:

\begin{equation}
h_i^{(l+1)}=\sigma\left(\sum_{j\in\mathcal{N}(i)}\alpha_{ij}^{(l)}\mathbf{W}^{(l)}h_j^{(l)}\right)
\label{eq:node_update}
\end{equation}

To further enhance the model's generalization capabilities under complex physical degradation scenarios, this paper introduces a multi-head attention mechanism (with the number of heads set to $K=2$). Multiple sets of independently computed features are concatenated or averaged at the end of the layer, supplemented by LayerNorm technology to accelerate the convergence of the deep network.

After 3 layers of GAT spatial message passing, the network successfully captures the local charge perturbation patterns induced by NBTI defect generation. Subsequently, this paper employs a global pooling operation (Global Mean \& Max Pooling) to compress the discrete node-level features into a graph-level representation that depicts the macroscopic state of the entire device. Finally, this graph-level representation is fed into a Multi-Layer Perceptron (MLP) decoder consisting of three fully connected layers (with 512, 256, and 50 nodes, respectively). The decoder directly outputs a transfer characteristic curve vector in the time domain containing 50 discrete bias points (i.e., the $I_D-V_G$ curve tensor), thereby accomplishing the end-to-end mapping from microscopic mesh physical quantities to macroscopic external characteristic curves.

\subsection{Dataset Generation and Model Training Setup}

\begin{figure}[pos=ht]
    \centering
    \includegraphics[width=\linewidth]{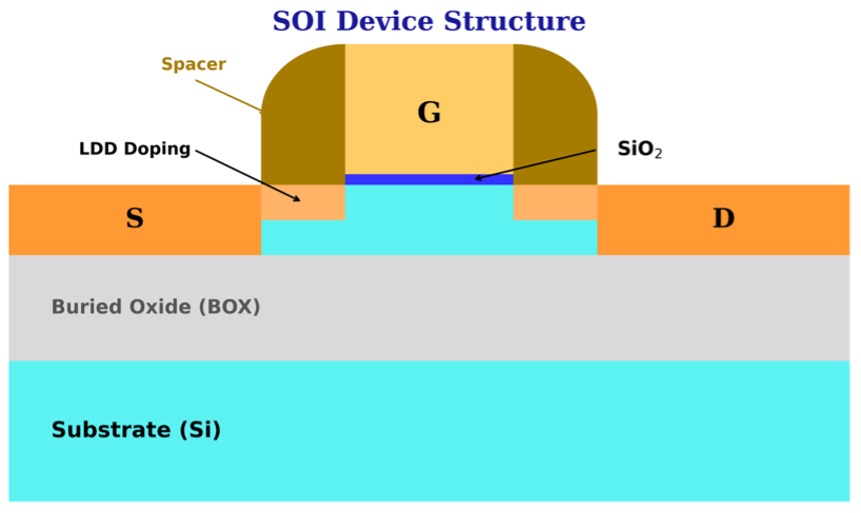}
    \caption{The cross-sectional structural model of the simulated PD-SOI PMOS device.}
    \label{fig:device_structure}
\end{figure}

To verify the effectiveness of the aging-aware graph encoding scheme and the GAT network, acquiring a high-quality dataset is imperative. To this end, we utilized TCAD tools for dataset generation. First, we constructed a structural model of a Partially Depleted Silicon-On-Insulator (PD-SOI) PMOS device, with the device structure illustrated in Fig. \ref{fig:device_structure}.

To endow the model with robust generalization capabilities, we performed spatial sampling on 9 key continuous physical parameters that dictate the device's electrical characteristics and aging behavior. These parameters encompass device geometric dimensions (e.g., gate length $L_{gate}$, oxide thickness $t_{ox}$), doping concentration profiles (source/drain $N_{sd}$, lightly doped drain $N_{ldd}$, channel $N_{ch}$), electro-thermal stress boundary conditions (ambient temperature $T$, stress gate bias $V_{g\text{\_stress}}$, stress drain bias $V_{d\text{\_stress}}$), and the precursor concentration ($conc$) of interface Si-H bonds that determines the degree of NBTI degradation in the two-stage model. To maximize the uniform coverage of the parameter space with a limited number of samples, this paper employs the Latin Hypercube Sampling (LHS) algorithm. Particularly, for doping parameters ($N_{sd}$, $N_{ldd}$, $N_{ch}$) and precursor concentration ($N_0$) that span multiple orders of magnitude, we perform LHS sampling within a logarithmic space to strictly adhere to their underlying physical distribution patterns.

\begin{table}[pos=ht]
\centering
\caption{Physical parameters and variation ranges for LHS sampling in dataset generation.}
\label{tab:physical_params}
\resizebox{\linewidth}{!}{%
\begin{tabular}{llcc}
\toprule
\textbf{Type} & \textbf{Parameter} & \textbf{Unit} & \textbf{Range} \\
\midrule
\multirow{2}{*}{Geometry} & Gate Length & $\mu$m & [0.09, 0.18] \\
 & Oxide Thickness & $\mu$m & [0.002, 0.005] \\
\midrule
\multirow{3}{*}{Doping} & Source/Drain Doping & cm$^{-3}$ & [$1 \times 10^{20}, 8 \times 10^{20}$] \\
 & LDD Doping & cm$^{-3}$ & [$1 \times 10^{18}, 8 \times 10^{18}$] \\
 & Channel Doping & cm$^{-3}$ & [$1 \times 10^{17}, 8 \times 10^{18}$] \\
\midrule
\multirow{3}{*}{Stress Condition} & Temperature & K & [250, 400] \\
 & Gate Bias & V & [-0.8, -1.6] \\
 & Drain Bias & V & [-0.8, -1.2] \\
\midrule
\multirow{2}{*}{Aging Parameter} & Precursor Concentration & cm$^{-3}$ & [$5 \times 10^{12}, 1 \times 10^{13}$] \\
 & Stress Time & s & \{0.1, 1, 10, 100, 500, 1000\} \\
\bottomrule
\end{tabular}%
}
\end{table}

Based on the aforementioned sampling strategy, we executed large-scale TCAD simulations across a vast physical design space. For each sampled device configuration, we extracted its transfer characteristic curves at 6 distinct aging stress time nodes ($t_{stress}$ = 0.1, 1, 10, 100, 500, 1000 s). Through the cross-combination of physical structures and aging parameters, we ultimately constructed a dataset comprising nearly 10,000 samples with varying structural configurations and degradation states. All sample data were strictly randomly partitioned into training, validation, and test sets according to the industry-standard ratio of 70\%, 20\%, and 10\%, respectively. Table \ref{tab:physical_params} details the physical parameters and their variation ranges utilized for dataset generation.

To validate the effectiveness of the proposed aging-aware encoding scheme, the detailed network architecture and hyperparameter configurations of the end-to-end degradation curve prediction model (Aging Predictor) are fully summarized in Table \ref{tab:hyperparameters}. Targeting the regression task of high-fidelity transfer characteristic curves, we deployed a deep RelGAT model with a total parameter count of approximately 1.863 M. The main body of this network consists of 3 dual-head graph attention layers (GAT Layers, Multi-head=2) and 4 linear fully connected layers. For graph-level feature extraction, the model adopts a combined Mean \& Max Pooling strategy to comprehensively capture the device's global attributes and local degradation extrema.

\begin{table}[pos=ht]
\centering
\caption{Detailed network architecture and hyperparameter configurations of the Aging Predictor.}
\label{tab:hyperparameters}
\resizebox{\linewidth}{!}{%
\begin{tabular}{ll}
\toprule
\textbf{Component / Hyperparameter} & \textbf{Aging Predictor (RelGAT)} \\
\midrule
Number of GAT layers & 3 \\
GAT layers (nodes) & 1 (104), 1 (232), 1 (512) \\
Number of Linear layers & 4 \\
Linear layers (nodes) & 1 (48), 1 (512), 1 (256), 1 (50) \\
Total layers & 7 \\
Pooling method & Mean \& Max \\
Normalization & LayerNorm \\
Residual connection & None \\
Multi-head attention & 2 \\
Total Parameters (M) & 1.863 \\
\bottomrule
\end{tabular}%
}
\end{table}

Regarding the optimization settings for model training, the entire network employs the Adam optimizer for parameter updates and introduces a customized piecewise linear learning rate decay strategy to enhance convergence in the later stages of training. Furthermore, considering that the current amplitude of the target $I_d-V_g$ curves spans several orders of magnitude, we uniformly transformed the macroscopic currents to a logarithmic scale for error calculation during regression training to ensure training stability. This targeted loss optimization strategy effectively circumvents the gradient imbalance issues caused by conventional mean squared error. The complete training process of the aforementioned model lasted for 300 Epochs.

\section{Results and Discussion}

\subsection{Full-Range I-V Prediction and Global Accuracy}

To evaluate the global prediction fidelity of the proposed Physics-Aware RelGAT surrogate model, we first analyzed the model's capability to predict static full-range transfer characteristics on an independent test set.

\begin{figure}[pos=ht]
    \centering
    \includegraphics[width=\linewidth]{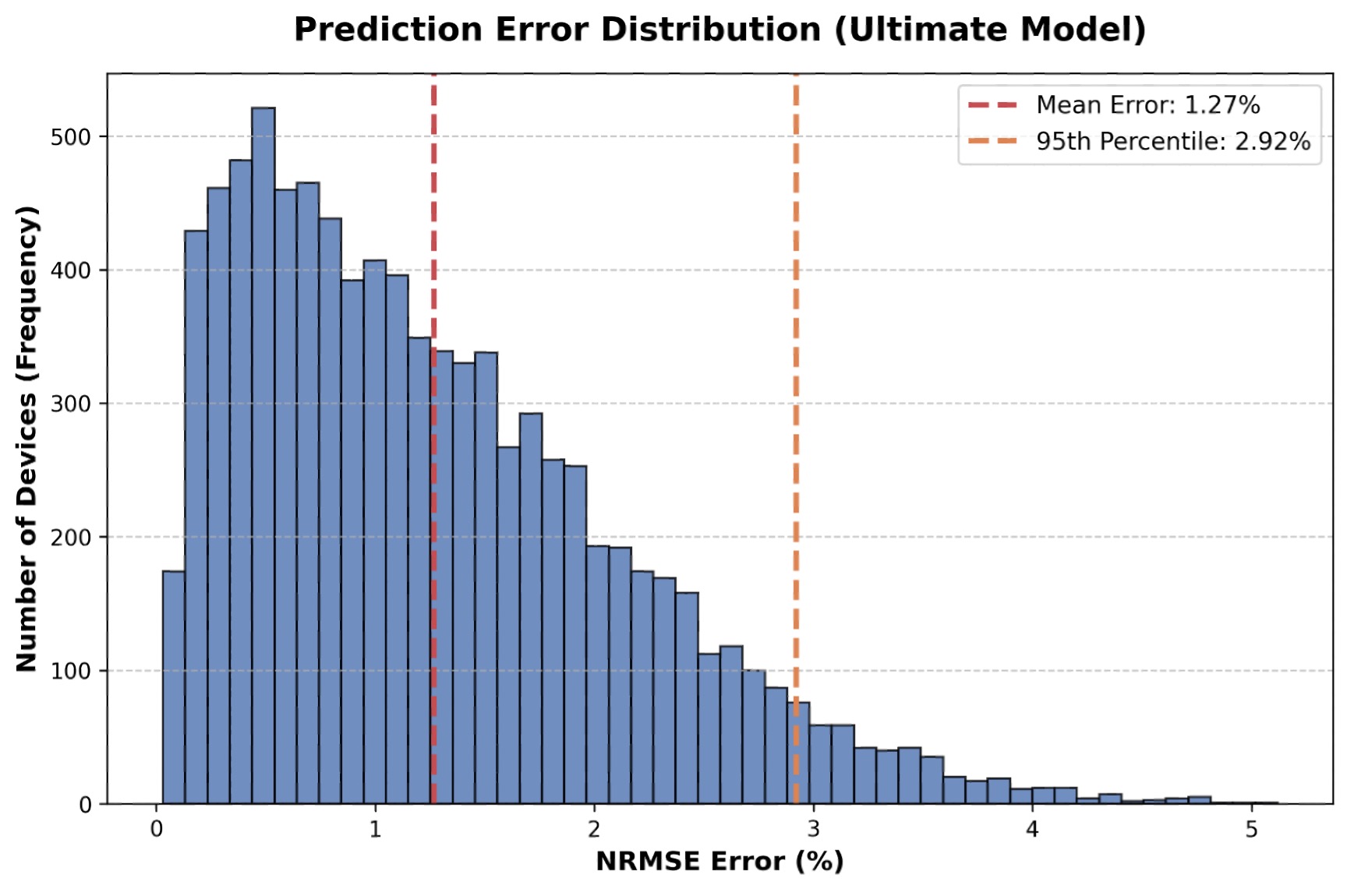}
    \caption{The NRMSE distribution histogram of the Physics-Aware RelGAT model on the independent test set.}
    \label{fig:error_dist}
\end{figure}

Fig. \ref{fig:error_dist} illustrates the distribution histogram of the Normalized Root Mean Square Error (NRMSE) across the entire test set. Benefiting from the introduction of the 45-dimensional deep physical features and the robust representation extraction capability of the residual graph attention network, the model exhibits excellent prediction accuracy within the highly complex parameter space. Statistical results indicate that the model's mean prediction error on the test set is merely 1.27\%, and the 95th percentile error is strictly constrained to within 2.92\%. This exceedingly low global error level demonstrates that the model has successfully captured the generalized physical laws across varying doping concentrations and device geometric dimensions.

\begin{figure}[pos=ht]
    \centering
    \includegraphics[width=\linewidth]{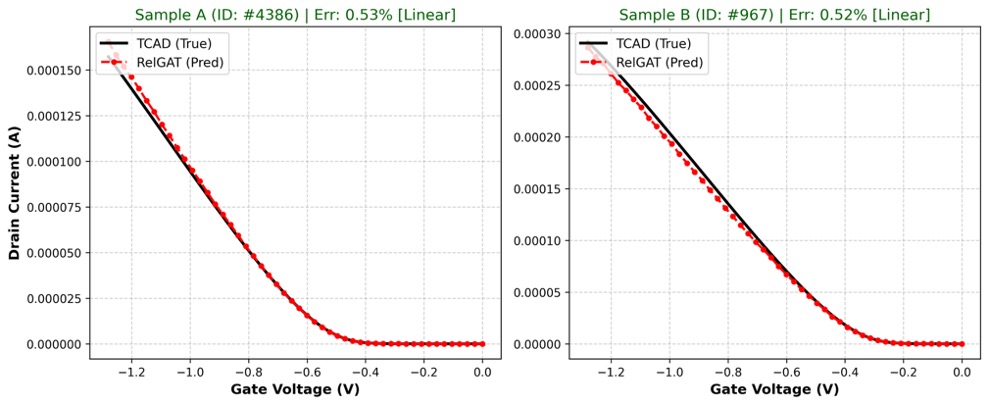}\\
    \vspace{0.3cm} 
    \includegraphics[width=\linewidth]{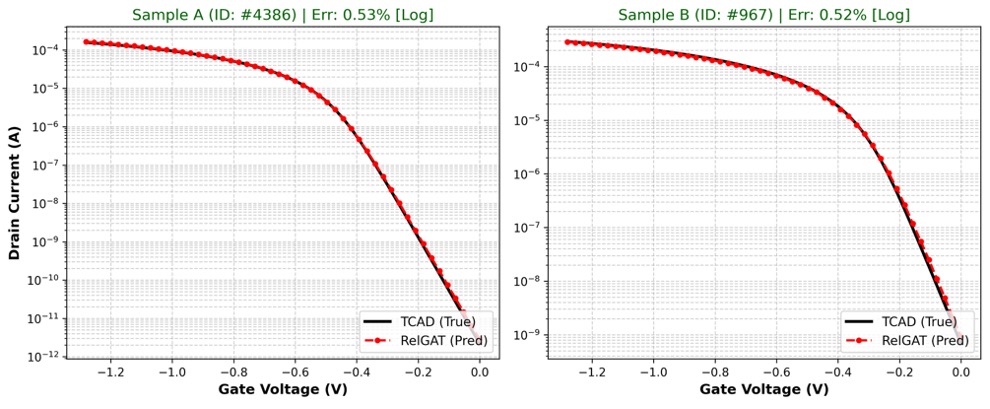}
    \caption{Comparison between TCAD simulated and RelGAT predicted transfer characteristic curves for two representative samples under (Top) linear scale and (Bottom) logarithmic scale.}
    \label{fig:iv_curves}
\end{figure}

The transfer characteristics of semiconductor devices span multiple orders of magnitude in current between the subthreshold region and the strong inversion region. Traditional surrogate models often struggle to balance both simultaneously. Fig. \ref{fig:iv_curves} randomly selects two representative typical samples from the test set and compares the ground-truth TCAD simulation curves with the RelGAT prediction curves under both logarithmic (log) and linear scales.

As shown in Fig. \ref{fig:iv_curves}, whether in the subthreshold region with extremely low gate voltage under the log scale, or in the ON-state high-current region under the linear scale, the prediction curves of RelGAT exhibit a near-perfect fit with the actual TCAD data. The model accurately captures the minute OFF-state leakage current ($I_{off}$) while maintaining exceedingly high fitting accuracy in the strong ON-state. This full-range, high-fidelity fitting spanning multiple orders of magnitude is attributed to the ``linear + log'' composite optimization strategy employed in the objective function. This strategy ensures that the network receives balanced and stable gradient guidance when processing both infinitesimal leakage currents and large saturation currents.

\subsection{Accurate Extraction of Key Physical Parameters ($V_{th}$ and $I_{ON}$)}

To meet the core requirements of subsequent circuit-level aging simulations (such as the extraction of aging SPICE models and the invocation of the OMI interface), we extracted two key parameters characterizing device aging degradation from the predicted $I_d-V_g$ curves: the threshold voltage ($V_{th}$) and the saturation ON-state current ($I_{ON}$).

In compact models geared towards advanced digital circuits, the accuracy of parameter extraction directly dictates the reliability of top-level circuit simulations. To this end, this paper employs the constant current method to extract $V_{th}$, and extracts the corresponding ON-state current $I_{ON}$ under strong inversion conditions.

\begin{figure}[pos=ht]
    \centering
    \includegraphics[width=\linewidth]{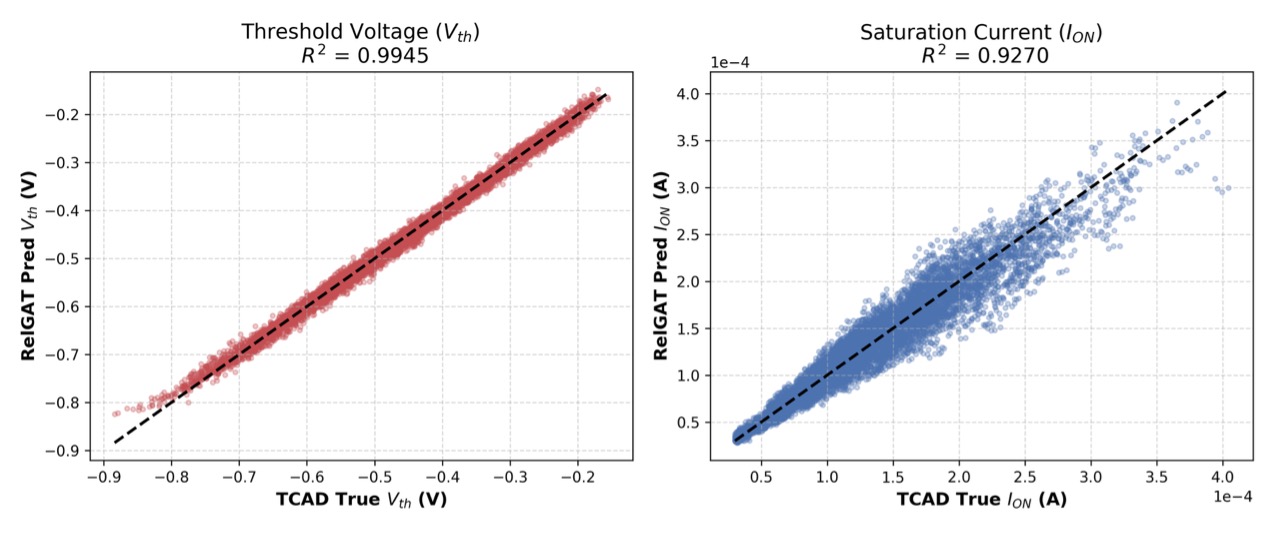}
    \caption{Scatter plots comparing the TCAD ground-truth values and RelGAT predictions for the extracted (Left) threshold voltage $V_{th}$ and (Right) ON-state current $I_{ON}$.}
    \label{fig:param_extraction}
\end{figure}

The scatter plots illustrated in Fig. \ref{fig:param_extraction} intuitively demonstrate the high degree of consistency between the TCAD ground-truth values and the RelGAT predictions. The predicted data points converge tightly around the ideal diagonal line ($y=x$). Quantitative statistical analysis reveals that the coefficient of determination ($R^2$ score) for the $V_{th}$ and $I_{ON}$ predictions reaches 0.9945 and 0.9270, respectively. This result strongly substantiates that the model not only exhibits excellent fitting on continuous curves but also provides highly fidelitous core discrete physical parameters for subsequent aging models and the reliability assessment of large-scale circuits.

\subsection{Prediction of Time-Dependent NBTI Degradation Curves}

To fully validate the capability of the RelGAT surrogate model in predicting dynamic parameter drifts, we extracted the family of transfer characteristic curves for a device with an identical structure at various aging times ($t_{stress}$ = 0.1 s, 10 s, 100 s, and 1000 s). 

As the stress time progresses, the device exhibits classic positive threshold voltage shifts and a drop in the saturation ON-state current ($I_{ON}$) due to the continuous accumulation of interface defects. Remarkably, the RelGAT model reproduces this complex time-domain evolutionary process with exceptionally high numerical fidelity. Quantitative evaluation demonstrates that the average log-scale Root Mean Square Error (Log-RMSE) for this family of curves across the four distinct time nodes is as low as 0.0291.

\begin{figure}[pos=ht]
    \centering
    \includegraphics[width=\linewidth]{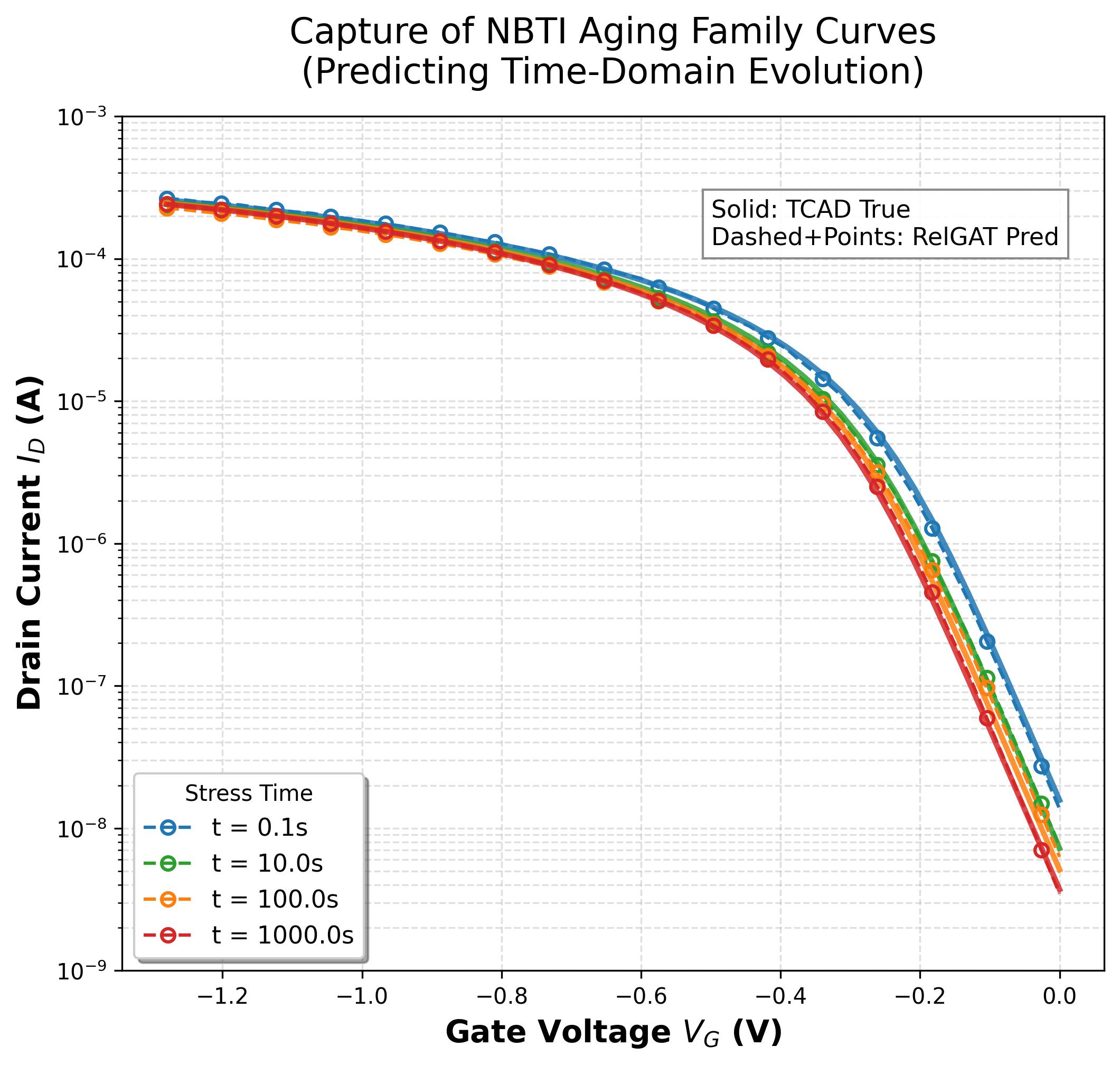}
    \caption{The family of transfer characteristic curves at different aging times, comparing the baseline TCAD physical simulations and RelGAT predictions across six orders of magnitude.}
    \label{fig:aging_curves}
\end{figure}

As illustrated in Fig. \ref{fig:aging_curves}, whether for the initial degradation at 10 s or the deep aging scenario at 1000 s, the predicted curves (dashed lines with open markers) and the baseline TCAD physical simulations (solid lines) exhibit a near-perfect overlap across a logarithmic current range spanning six orders of magnitude. This result powerfully demonstrates that RelGAT can successfully decouple the physical correlations between static process parameters and dynamic aging time from within its internal high-dimensional (45-dimensional) feature space. Consequently, within an exceedingly short inference time, it offers a highly trustworthy underlying curve verification solution for SPICE compact aging models.

\subsection{Computational Efficiency for Fast DTCO}

To demonstrate the computational advantages of the proposed model, we conducted a performance benchmarking test comparing the traditional TCAD simulator with the trained surrogate model. The benchmarking task involved generating 1,000 independent sets of multi-bias NBTI transfer characteristic curves. On a high-performance workstation equipped with dual-socket server processors (Intel Xeon Gold 6248R, 96 physical cores), the TCAD physical simulator, even under parallel computational scheduling, consumed up to 15 hours and 3 minutes (54,180 seconds) to complete the iterative solving for all simulation tasks.

In stark contrast, the proposed aging-aware RelGAT model completely subverts the traditional numerical solving paradigm. Through the message passing mechanism of graph neural networks, RelGAT reduces the iterative solving of complex mathematical and physical equations to a single forward inference of static matrices. For the identical set of 1,000 test cases, RelGAT required merely 3.17 seconds to generate the transfer characteristic curves with a stringent average error of 1.27\%. The average inference time for a single curve is exceptionally low, reaching down to 3.17 milliseconds (ms).

These results indicate that, when generating aging curves of equivalent quality, the proposed model achieves a staggering acceleration factor of 17,091$\times$ (approximately 17,000 times) compared to conventional TCAD. This extraordinary efficiency enhancement, spanning four orders of magnitude, significantly alleviates the computational bottlenecks inherent in traditional physical simulations. Consequently, it drastically shortens the iteration cycle for aging statistical analysis and process optimization, providing a highly practical and rapid evaluation tool to effectively drive the reliability-aware DTCO flow.

\section{Conclusion}

This paper presents and validates an NBTI aging evaluation framework based on physics-aware relational graph attention networks (RelGAT). By losslessly mapping unstructured TCAD meshes into graphs and integrating a 45-dimensional device encoding scheme that encompasses aging conditions such as microscopic interface trap distributions and macroscopic electro-thermal stresses, this framework imposes stronger physical constraints compared to traditional black-box models, thereby achieving an end-to-end mapping from underlying physical quantities to device degradation characteristics. 

Experimental results demonstrate that the RelGAT model aligns exceptionally well with the TCAD simulation baseline across the full bias sweep range, yielding a minute average prediction error of only 1.27\% and accurately capturing the degradation characteristics of key reliability indicators ($V_{th}$, $I_{dsat}$). Furthermore, this approach transforms the iterative solving of complex non-linear partial differential equations into efficient tensor forward inference, achieving an approximately 17,000$\times$ simulation speedup, with the inference time for a single degradation curve dropping to as low as 3.17 milliseconds. 

In conclusion, this study not only demonstrates the efficacy of graph neural networks in modeling complex physical fields but also provides a practical acceleration scheme that seamlessly balances physical fidelity with computational efficiency for reliability-aware DTCO workflows in advanced technology nodes.

\bibliographystyle{elsarticle-num}
\bibliography{cas-refs}

@inproceedings{Mukhopadhyay2023,
  author    = {Mukhopadhyay, S. and others},
  title     = {A Unified Aging Model Framework Capturing Device to Circuit Degradation for Advance Technology Nodes},
  booktitle = {2023 IEEE International Reliability Physics Symposium (IRPS)},
  year      = {2023},
  pages     = {1-6},
  address   = {Monterey, CA, USA}
}

@inproceedings{Chatterjee2023,
  author    = {Chatterjee, Payel and others},
  title     = {A Device to Circuit Framework for {NBTI}},
  booktitle = {2023 IEEE International Integrated Reliability Workshop (IIRW)},
  year      = {2023},
  publisher = {IEEE}
}

@inproceedings{Kelleher2022,
  author    = {Kelleher, A. B.},
  title     = {Celebrating 75 years of the transistor: A look at the evolution of Moore's Law innovation},
  booktitle = {2022 IEEE International Electron Devices Meeting (IEDM)},
  year      = {2022},
  address   = {San Francisco, CA, USA}
}

@inproceedings{Moroz2020,
  author    = {Moroz, V. and others},
  title     = {{DTCO} launches Moore's law over the feature scaling wall},
  booktitle = {2020 IEEE International Electron Devices Meeting (IEDM)},
  year      = {2020},
  publisher = {IEEE}
}

@inproceedings{Novkin2023,
  author    = {Novkin, R. and Thomann, S. and Amrouch, H.},
  title     = {{ML-TCAD}: Perspectives and Challenges on Accelerating Transistor Modeling using {ML}},
  booktitle = {2023 ACM/IEEE 5th Workshop on Machine Learning for CAD (MLCAD)},
  year      = {2023},
  pages     = {1-6},
  address   = {Snowbird, UT, USA}
}

@article{Xue2023,
  author    = {Xue, Yongkang and others},
  title     = {On the understanding of pMOS {NBTI} degradation in advance nodes: Characterization, modeling, and exploration on the physical origin of defects},
  journal   = {IEEE Transactions on Electron Devices},
  volume    = {70},
  number    = {9},
  pages     = {4518-4524},
  year      = {2023}
}

@inproceedings{Zhao2019,
  author    = {Zhao, Y. and others},
  title     = {A Unified Physical {BTI} Compact Model in Variability-Aware {DTCO} Flow: Device Characterization and Circuit Evaluation on Reliability of Scaling Technology Nodes},
  booktitle = {Symposium on VLSI Technology},
  year      = {2019},
  address   = {Kyoto, Japan}
}

@article{Tariq2025,
  author    = {Tariq, A. and others},
  title     = {Optimizing Silicon {MOSFETs}: The Impact of {DTCO} and Machine Learning Techniques},
  journal   = {Electronics},
  volume    = {14},
  number    = {3},
  year      = {2025}
}

@inproceedings{Mohamed2025,
  author    = {Mohamed, Tarek and Amrouch, Hussam},
  title     = {Accelerating Reliability Analysis for Aging and Self-Heating using Machine Learning},
  booktitle = {2025 26th International Symposium on Quality Electronic Design (ISQED)},
  year      = {2025},
  publisher = {IEEE}
}

@inproceedings{Han2021_conf,
  author    = {Han, S.-C. and Choi, J. and Hong, S.-M.},
  title     = {Acceleration of Three-Dimensional Device Simulation with the {3D} Convolutional Neural Network},
  booktitle = {2021 International Conference on Simulation of Semiconductor Processes and Devices (SISPAD)},
  year      = {2021},
  pages     = {188-191},
  address   = {Dallas, TX, USA}
}

@article{Han2021_jour,
  author    = {Han, Seung-Cheol and Choi, Jonghyun and Hong, Sung-Min},
  title     = {Acceleration of semiconductor device simulation with approximate solutions predicted by trained neural networks},
  journal   = {IEEE Transactions on Electron Devices},
  volume    = {68},
  number    = {11},
  pages     = {5483-5489},
  year      = {2021}
}

@article{Wei2022_tcad,
  author    = {Wei, Jiahao and others},
  title     = {A new compact {MOSFET} model based on artificial neural network with unique data preprocessing and sampling techniques},
  journal   = {IEEE Transactions on Computer-Aided Design of Integrated Circuits and Systems},
  volume    = {42},
  number    = {4},
  pages     = {1250-1254},
  year      = {2022}
}

@article{Wei2022_integ,
  author    = {Wei, JiaHao and others},
  title     = {Modeling of {CMOS} transistors from 0.18 {$\mu$m} process by artificial neural network},
  journal   = {Integration},
  volume    = {87},
  pages     = {11-15},
  year      = {2022}
}

@article{Jang2023,
  author    = {Jang, W. and others},
  title     = {{TCAD} Device Simulation With Graph Neural Network},
  journal   = {IEEE Electron Device Letters},
  volume    = {44},
  number    = {8},
  pages     = {1368-1371},
  month     = {Aug},
  year      = {2023}
}

@article{Fan2025,
  author    = {Fan, G. and others},
  title     = {Graph Attention Network-Based Unified {TCAD} Modeling Enabling Fast Design Technology Co-Optimization Through Transfer Learning},
  journal   = {IEEE Transactions on Electron Devices},
  volume    = {72},
  number    = {1},
  pages     = {474-480},
  month     = {Jan},
  year      = {2025}
}

@article{Stathis2006,
  author    = {Stathis, J. H. and Zafar, S.},
  title     = {The negative bias temperature instability in {MOS} devices: A review},
  journal   = {Microelectronics Reliability},
  volume    = {46},
  number    = {2-4},
  pages     = {270-286},
  year      = {2006}
}

@inproceedings{Grasser2009,
  author    = {Grasser, T. and others},
  title     = {A Two-Stage Model for Negative Bias Temperature Instability},
  booktitle = {IEEE International Reliability Physics Symposium (IRPS)},
  year      = {2009},
  pages     = {33-44},
  month     = {April},
  address   = {Montréal, Québec, Canada}
}

@inproceedings{Goes2009,
  author    = {Goes, W. and others},
  title     = {A Model for Switching Traps in Amorphous Oxides},
  booktitle = {International Conference on Simulation of Semiconductor Processes and Devices (SISPAD)},
  year      = {2009},
  pages     = {159-162},
  month     = {September},
  address   = {San Diego, CA, USA}
}

\end{document}